\documentclass[12pt,preprint]{aastex}
\usepackage{graphicx}
\usepackage{fancyheadings}
\usepackage{color}
\usepackage{natbib}

\begin{document}

\title{Probing the Nature of the Vela X Cocoon}

\author{Stephanie M.  LaMassa\altaffilmark{1}, Patrick O. Slane\altaffilmark{2}, \& Okkie C. de Jager\altaffilmark{3,4}}

\altaffiltext{1}{The Johns
  Hopkins University} 

\altaffiltext{2}{Harvard-Smithsonian Center for
  Astrophysics} 

\altaffiltext{3}{Unit for Space Physics, North-West University}

\altaffiltext{4}{South African 
  Department of Science \& Technology and National Research Foundation
  Research Chair: Astrophysics \& Space Science}

\begin{abstract}

Vela~X is a pulsar wind nebula (PWN) associated with the active
pulsar B0833-45 and contained within the Vela supernova remnant
(SNR). A collimated X-ray filament (``cocoon'') extends south-southwest
from the pulsar to the center of Vela~X. VLA observations uncovered
radio emission coincident with the eastern edge of the cocoon and
H.E.S.S. has detected TeV $\gamma$-ray emission from this region as
well. Using XMM-\textit{Newton} archival data, covering the southern
portion of this feature, we analyze the X-ray properties of the
cocoon. The X-ray data are best fit by an absorbed nonequilibrium
plasma model with a powerlaw component. Our analysis of the thermal
emission shows enhanced abundances of O, Ne, and Mg within the
cocoon, indicating the presence of ejecta-rich material from the
propagation of the SNR reverse shock, consistent with Vela~X being
a disrupted PWN. We investigate the physical processes that excite
the electrons in the PWN to emit in the radio, X-ray and $\gamma$-ray
bands.  The radio and non-thermal X-ray emission can be explained
by synchrotron emission. We model the $\gamma$-ray emission by Inverse
Compton scattering of electrons off of cosmic microwave background
(CMB) photons. We use a 3-component broken power law to model the
synchrotron emission, finding an intrinsic break in the 
electron
spectrum at $\sim5 \times 10^{6}$ keV and a cooling break at $\sim$ 5.5
$\times 10^{10}$ keV. This cooling break along with a magnetic field
strength of 5 $\times 10^{-6}$ G indicate that the synchrotron break
occurs at $\sim$1 keV.

\end{abstract}


\section{Introduction}

The Vela supernova remnant (SNR), at a distance of $\sim 290$ pc
\citep{Dod} is the closest SNR to contain an active pulsar, (PSR)
B0833-45. The Vela pulsar has a period (\textit{P}) of 89 ms and a
period derivative (\textit{\.{P}}) of 1.25 x 10 $^{-13}$s s$^{-1}$,
implying an age of 11,000 years and a spin-down luminosity
(\textit{\.{E}}) of 7$\times 10^{36}$ erg s$^{-1}$. The remnant
spans about 8$^\circ$ in diameter and contains regions of nonthermal
emission in the radio, X-ray and $\gamma$-ray bands, including Vela
X, a pulsar wind nebula (PWN) spanning a region of 2$^\circ$ $\times$
3$^\circ$ south-southwest of the pulsar. The center of Vela~X is
offset by about 40' from the pulsar and emits in the radio, X-ray,
and $\gamma$-ray bands (Frail et al. 1997; Aharonian et al. 2006).

\begin{figure}[f]
\plotone{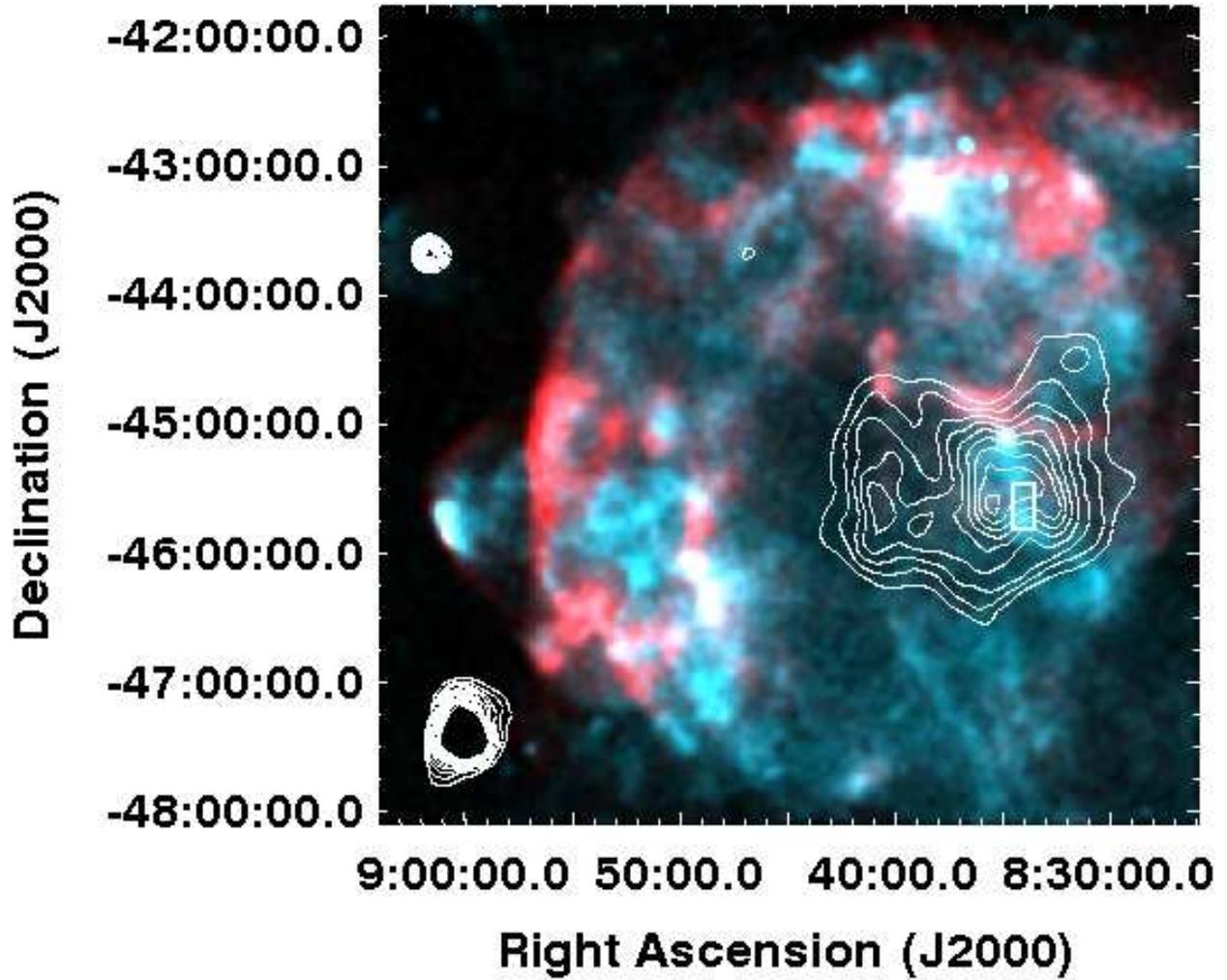}
\caption{\label{rosat}Composite \textit{ROSAT} RASS image of the Vela SNR with
  radio contours overlaid. The red band is the lower energy range,
  from 0.1 - 0.4 keV and the green and blue bands are the higher energy range,
  from 0.5 - 2.0 keV. All bands were smoothed using a Gaussian kernel
  of 75 arcseconds. The white box indicates the XMM-\textit{Newton} region we analyzed.}
\end{figure}

\textit{ROSAT} observations of the Vela SNR (Figure~\ref{rosat})
reveal a collimated X-ray filament seen predominantly at higher
energies extending from the pulsar to the center of Vela~X, about
45' in length. Markwardt and \"{O}gelman (hereafter M\"{O}) interpreted
the \textit{ROSAT} feature as an X-ray jet along which the pulsar
loses energy \citep{MO95}. 
 The ``jet head'', or interaction
between the jet and SNR shell, was later observed with \textit{ASCA}
by M\"{O}, and they found the spectrum of this region to be best
fit by a two component model, a lower temperature thermal plasma model
and a higher energy component, that could be fit by either a higher
temperature plasma model or power-law emission \citep{MO97}.

An alternate explanation for the X-ray filament was proposed by
Gvaramadze (1999) who suggested that this feature represents a
projection effect of a Rayleigh-Taylor instability, which is caused
by the impact of the SN ejecta with the wind-driven shell from the
progenitor. Subsequent \textit{Chandra} observations revealed an
X-ray jet situated along the northwest-southeast pulsar spin axis,
coinciding with the pulsar proper motion and not overlapping the
M\"{O} ``jet'' \citep{Pavlov}.

Though this collimated X-ray feature (hereafter ``cocoon'') is no
longer considered a ``jet,'' it warrants further investigation.
VLA observations reveal a bright radio filament coincident with the
eastern edge of this filament \citep{Frail}.  H.E.S.S. observed TeV
$\gamma$-rays emanating from the \textit{ROSAT} cocoon; Aharonian
et al. 2006 use Inverse Compton scattering to fit this H.E.S.S.
spectral energy distribution (SED). Horns et al. 2006 attribute the
$\gamma$-ray emission from the cocoon to hadronic processes, which
hinges on a thermal plasma density of 0.6 cm$^{-3}$, consistent
with the previous findings of  M\"{O} 1995 .

Here we report on the analysis of XMM-\textit{Newton} archival data
from the interaction region of this filament with the SNR shell. In
\S2 we summarize the observations and data reduction techniques
applied to the associated spectra. In \S3 we describe our modeling
of the X-ray spectrum, where we investigate the thermal and nonthermal
emission observed from the cocoon. We investigate the broadband
spectrum from this region in \S4, using the results of our X-ray
analysis along with spectral information from the radio and very
high energy $\gamma$-ray bands. We consider these results in the
context of a model in which the reverse shock from the Vela SNR has
interacted with the PWN (Blondin, Chevalier, \& Frierson 2001). We
discuss our conclusions in \S5.

\section{Observations and Data Reduction}

The ``Vela jet head'' was observed by XMM-\textit{Newton} on May
2, 2005 for 22 ks with the MOS1 and MOS2 detectors and for 20 ks
with the PN detector in Prime Full Window mode with the medium
filter (PI Gallant, ObsID 0094630101).
 Data reduction was
performed with the XMM Science Analysis System (SAS) to filter the
data and remove flares. The good science time remaining after this
processing is as follows: 22 ks for MOS1, 21.6 ks for MOS2, and
17.6 ks for PN.

\begin{figure}[f]
\plotone{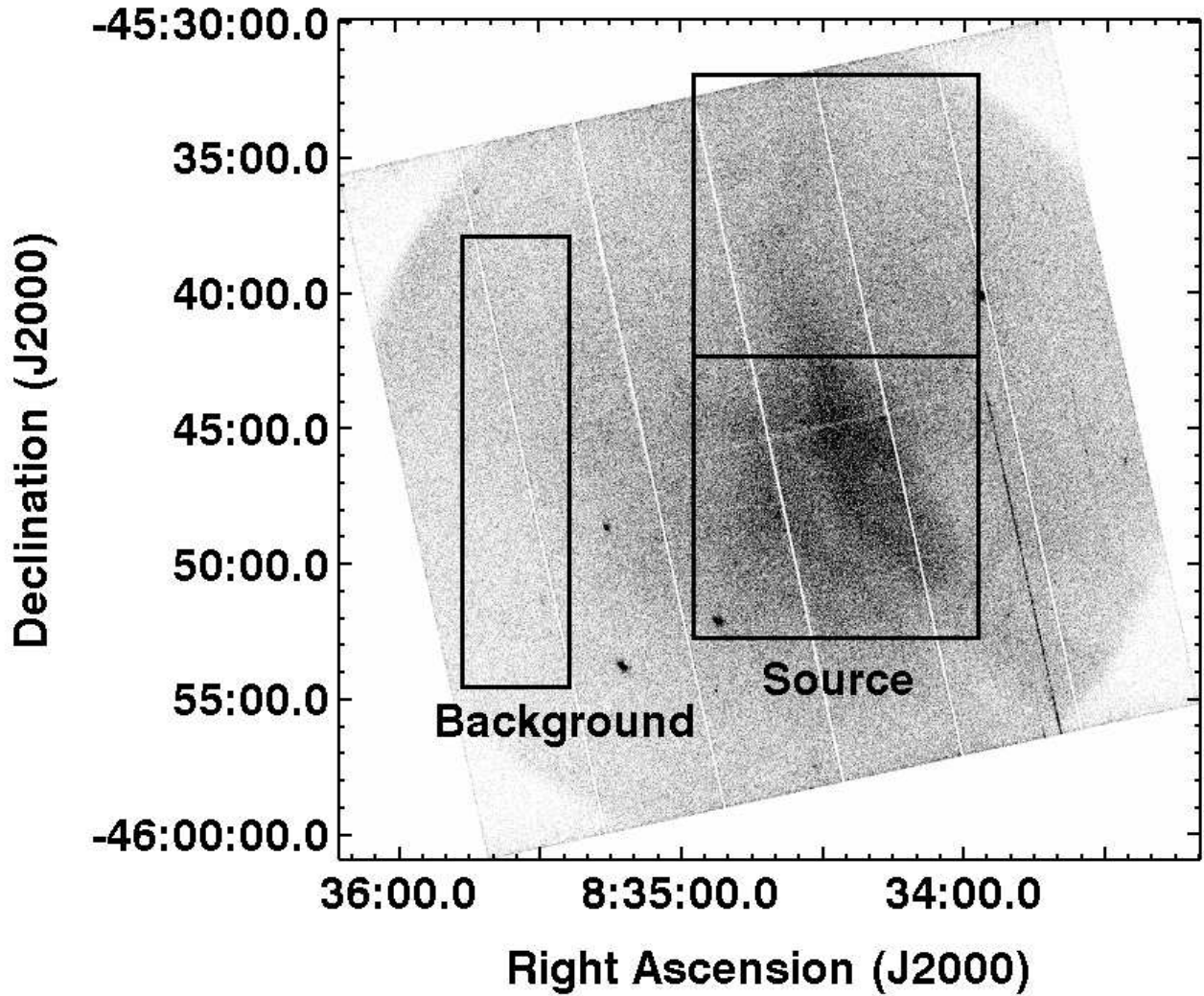}
\caption{\label{pn}XMM-\textit{Newton} image from the PN detector
  of the southern portion of the Vela~X cocoon, filtered in energy
  from 0.2 to 6.6 keV.  The source and background regions are
  indicated by the black boxes, where both the northern and southern
  source regions indicate the regions used when investigating the
  northern and southern portions of the cocoon separately. The
  northern portion covers the region about 22.6' to 31.8' from the
  pulsar and the southern portion extends from about 31.8' to 42.1'
  from the pulsar. The image was
  smoothed with a 
  gaussian kernel of 9.6''.}
\end{figure}

The XMM field of view covers just the southernmost portion of the
cocoon detected by \textit{ROSAT}. The field of view for the PN
detector is shown in Figure~\ref{pn}, with the source and background
regions we analyzed overlaid.  Spectra were extracted from roughly
the same source and background regions among all 3 detectors, using
the \textit{xmmselect} gui. The response files were generated with
the \textit{rmfgen} and \textit{arfgen} SAS tools.

We extracted background spectra from a region within the
the SNR and the 2$^\circ$ x 3$^\circ$ radio PWN, but outside the cocoon (Figure 2).
For comparison, we extracted a second set of background spectra
from a region along the northwestern shell of the SNR using archival
data from observations performed on 9 November 2001 (PI Turner, ObsID
0112690101). The latter region is centered at RA = 08:48:33 and Dec =
-42:23:45, with a radius of $1.8^\prime$.

We applied the SAS tool evigweight to all spectra to correct for
vignetting.  To account for the unvignetted internal background,
for both the source and background regions we subtracted similarly
processed spectra taken from available blank sky fields \footnote
{Blank sky files were taking in Prime Full Frame mode using the
medium filter (M1.M.FF.EVLIRP.FIT, M2.M.FF.EVLIRP.FIT, and
PN.M.FF.EVLIRA.FIT).}, each weighted by a factor determined by the
ratio of high-energy counts between the Vela and blank-sky spectra
(see Arnaud et al. 2002).

The source region was also separated into upper and lower sections,
as seen in Figure~\ref{pn}. Spectra were extracted separately from
these areas to search for possible spectral changes with increasing
distance from the pulsar.

\section{Spectral Analysis}

\begin{figure}[f]
\plotone{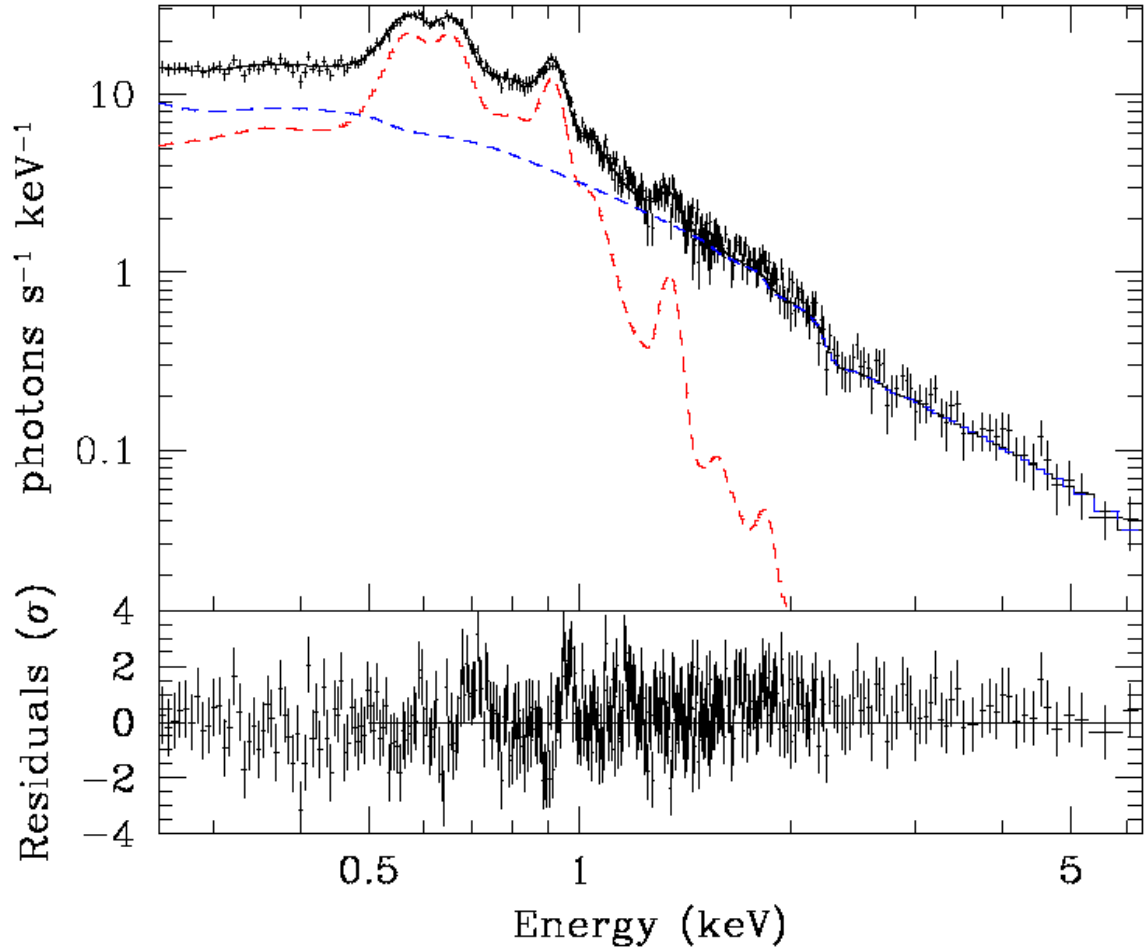}
\caption{\label{spectrum}Epic PN spectrum from Vela X cocoon. The solid curve represents the best-fit model
summarized in Table 1. Individual model components are shown in red (thermal) and blue (nonthermal).}
\end{figure}

We modeled the spectra in XSPEC and found that a two-component model
was required to fit the entire source region, consistent with the
findings of M\"{O} in 1997.  Using an adjacent background region,
the data for the full region were best fit by an absorbed (N$_H$
$\sim$ 1.6 $\times$ 10$^{20}$) nonequilibrium plasma model with a
powerlaw component to accommodate the hard emission, as shown in
Figure~\ref{spectrum}. A very high temperature thermal component proposed by M\"{O} (1997)
can reproduce the observed hard
emission in the XMM band, but given the observed emission from Vela~X at energies
in excess of 100 keV \citep{Mang}, it appears clear that this component
is nonthermal. The results of the
plasma model with a powerlaw component are presented in the first
column of Table 1. The modeling shows enhanced abundances of oxygen,
neon, and magnesium, higher than abundances seen in the SNR as a
whole, but consistent with abundances represented in ejecta fragments,
most notably Vela Region D \citep{Miceli}. The temperature of
$\sim$0.5 keV is also higher than that for most of the remnant,
which ranges between 0.12 and 0.18 keV \citep{Lu}, but is consistent
with temperatures of shocked ejecta fragments \citep{Asch}. The
hard component is best fit by a photon index of $\Gamma$ of 2.30
$\pm$ 0.04, consistent with synchrotron emission from energetic particles 
injected by the pulsar.

We compare these results using a background region from 
within the northwestern rim of the SNR shell \citep{Miceli}.  
By subtracting this background region, the source spectra retains
contribution from the PWN, but contributions from the background
and SNR are removed.  The best-fit parameters from this fit
are shown in the second column of Table 1. The abundances are still
elevated after subtracting out SNR emission, suggesting that high
abundances are intrinsic to this portion of the
cocoon.

Spectra were also extracted separately from the upper and lower
regions of the portion of the cocoon covered in this pointing, as
shown in Figure~\ref{pn}, to investigate whether the steepness of
the power law spectrum changes with distance from the pulsar.  The
northern portion covers the region about 22.6' to 31.8'($\sim$2.0
pc) from the  pulsar and the southern portion extends from about
31.8' to 42.1' ($\sim$3.7 pc) from the pulsar. To facilitate fitting
the spectra, $N_H$ was set to our best fit value of 1.6$\times 10^{20}$ and 
$\tau$ was set to 4.8$\times 10^{10}$. A variation in the spectral 
index was found, with 
$\Gamma_{upper}=2.13 \pm 0.03$ and $\Gamma_{lower}=1.90 \pm 0.02$. Setting
$N_H$ to the the value in the literature (2.26$\times 10^{20}$, M\"{O} 1997) shows
a similar trend, with $\Gamma_{upper}$ = 2.22$\pm$0.02 and $\Gamma_{lower}$ = 1.99$\pm$0.01. 
The $\Gamma$ values above
indicate that the power-law spectrum does change with distance from
the pulsar, but in the opposite manner expected, flattening at
larger radii rather than steepening. Further
observations along the length of the cocoon are needed to fully
test whether this trend exists along the full filament or if this
is a local phenomenon related to turbulent disruption of the PWN
where crushing of the cocoon against the PWN could cause adiabatic
heating that hardens the spectrum.

\section{Modeling}

We have modeled the radio and X-ray emission as synchrotron radiation
from energetic electrons within the cocoon, 
and the $\gamma$-ray emission 
as
Inverse Compton emission from the
scattering of the cosmic microwave background radiation off of the
same electron population. The domain
between the radio and X-ray energies is currently unexplored so we
have interpolated between these two bandpasses. 
In order to reproduce the spectral index derived from the X-ray
described above, the measured spectral indices in the radio band (with 
$\alpha_R$ ranging from $\sim 0$ to 0.38, where $S_\nu \propto \nu^{-\alpha_R}$),
and also the observed spectral shape of the TeV emission, we require an electron
spectrum with multiple slopes.
Treating the electron spectrum as a power law with a single break can adequately describe
the X-ray and $\gamma$-ray spectra, but either under-predicts the radio emission by a significant
factor, or requires a particle spectrum that leads to $\alpha_r \sim 0.7$, much steeper
than the measured values.
To fit the entire spectrum,
we thus consider a three-component power law:

\[ \frac{dN}{dE} = \left\{ \begin{array}
  {r@{\quad:\quad}l}
  A_{1}E^{-\sigma_{1}} & E \le E_{1} \\ 
  A_{2}E^{-\sigma_{2}} & E_{1} < E \le E_{2} \\
  A_{3}E^{-\sigma_{3}} & E > E_{2} \\

\end{array}\right. \]
A$_{1}$, A$_{2}$, and A$_{3}$ are the normalization factors and
$\sigma_{1}$, $\sigma_{2}$, and $\sigma_{3}$ are the indices of the
electron distribution within each energy range. The photon spectra
produced by the synchrotron and IC emission in ${\nu}F_{\nu}$ format are:

\[{\nu}F_{\nu}(synch) = \frac{V_{E}}{4{\pi}D^{2}_{pc}}{\nu}P_{tot}({\nu},E),
\]

\[{\nu}F_{\nu}(IC) = \frac{V_{E}}{4{\pi}D^{2}_{pc}}E^{2}\frac{dn_{\gamma}({\epsilon}_{\gamma})}{dt},\]
where $V_{E}\sim 10 ^{55}{\rm\ cm}^3$ is the electron emission volume
assuming a cylindrical volume with a radius of $10^\prime.44$ and
length of $19^\prime.5$; here we have assumed a distance $D_{pc} = 290$~pc
\citep{Mang}.  $P_{tot}({\nu},E)$ 
and $dn_{\gamma}({\epsilon}_{\gamma})/dt$ are synchrotron and IC
emissivities \\ \citep{Lanz}.

\begin{figure}[f]
\plotone{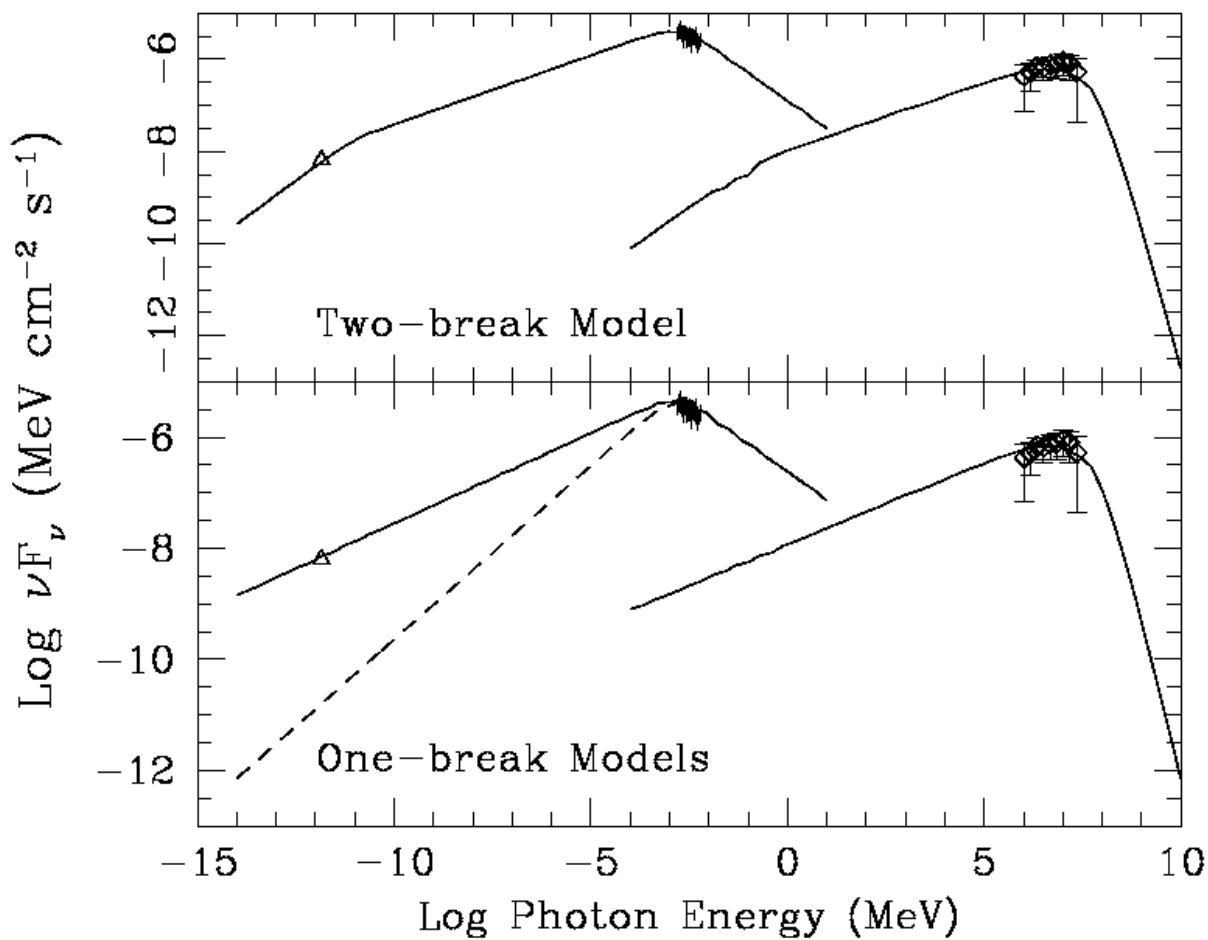}
\caption{\label{model}Broadband modeling of cocoon emission. The upper panel shows a power law electron
spectrum with two spectral breaks. The lower panel shows a power law spectrum with a single break, using
a low-energy spectral index that either reproduces the observed radio flux (solid curve) or matches the 
measured radio spectral index (dashed curve). The data points from the VLA and H.E.S.S. were scaled to match
the size of the region in the XMM observation.}
\end{figure}

The VLA, H.E.S.S., and non-thermal components of the XMM spectra were
plotted along with the flux 
model cited above. The VLA \citep{Frail} and H.E.S.S. data
points were scaled to coincide
with the XMM region size. The magnetic field strength, normalization, spectral index
values, and break energy values were varied until we
obtained a model that adequately described the data points, as shown
in Figure~\ref{model}; A$_{2}$ 
and A$_{3}$ were calculated by fixing A$_{1}$ and requiring the flux
to match at the endpoints of the broken power-law segments. The best-fit
parameters from this fit are shown in Table 2. 

A power law spectrum with multiple breaks has been used to describe
the emission from a number of PWNe, most notably the Crab Nebula.
The first break in the spectrum of the Vela X cocoon, at $5\times 10^{6}$
keV, most likely corresponds to an intrinsic break. Several factors can
contribute to this break; for the Crab nebula in particular, it
has been posited that this break results from two distinct electron
populations or from an elevated value of the
magnetization parameter \citep{Volpi}. It has been suggested by
de Jager (2007) that the radio-emitting and X-ray-emitting electrons
from the entire PWN form distinct populations. We note that our constraints
on both the maximum energy and the overall normalization of this low-energy
component are somewhat weak, and do not rule out this possibility for
the southern cocoon region.

The second break in the spectrum at $5.5\times 10^{10}$ keV is most
likely a cooling break related to the synchrotron lifetime of the
electrons injected from the pulsar. For our analysis of this segment
of the Vela PWN, the magnetic field (5$\times10^{-6}$ G) and cutoff
energy (5.5$\times 10^{10}$ keV) corresponds to a critical frequency
of 2.4$\times 10^{17}$ Hz, or $\sim$1 keV, indicating that the
synchrotron break occurs in the X-ray band. This magnetic field
strength and synchrotron break are consistent with the values derived
by de Jager, Slane \& LaMassa (2008) which used \textit{INTEGRAL},
BeppoSAX, and H.E.S.S. observations of the full Vela X region to
constrain a model of a postshock injection spectrum. In this case,
the electron spectra are produced by solving the time dependent
transport equation with synchrotron losses.
Integration of the non-thermal electron spectrum from this region of
the cocoon gives an energy
of $\sim 7.7 \times10^{45}$~erg, which is about 0.3\% of the
energy input from the pulsar ($E \approx$ \textit{\.{E}} $\times \tau_c \sim 2.31 \times10^{48}$
erg, where $\tau_c$ is the characteristic age of the pulsar).

\section{Discussion and Conclusions}

If the ambient medium into which a SNR expands is inhomogeneous,
simulations show that the SNR will expand asymmetrically, resulting
in a reverse shock that subsequently crushes the PWN asymmetrically
\citep{Blondin}.  The Vela radio nebula extends southward of the
pulsar, possibly suggesting that the northern portion has been
disturbed by such an interaction with the SNR reverse shock.
In such a scenario, simulations show that ejecta-rich material can
be turbulently mixed into the PWN from the reverse shock interaction
\citep{Blondin}.
This is consistent with our observation of 
enhanced abundances of O, Ne, and Mg in the cocoon, 
similar to the observed abundances of ejecta fragments
in Vela (e.g. Vela Region D, Miceli et al. 2005),
providing support for the disrupted PWN interpretation.

Our modeling of the non-thermal emission from this segment of the cocoon
indicates that a leptonic model adequately describes the data, with
the radio and X-ray emission resulting from synchrotron radiation and
the $\gamma$-ray emission arising from inverse-Compton scattering.
We need a 3-component broken power law model to adequately fit the data, unless we 
ignore the radio flux and spectral index values for Vela X.
The high energy break, at 5.5$\times 10^{10}$ keV, and derived magnetic
field of 5$\times 10^{-6}$ G, correspond to a break in the synchrotron
spectrum at  $\sim$1
keV and are consistent with the findings of de
Jager, Slane \& LaMassa (2008) which obtained
obtain the same field strength and synchrotron break using
a different approach. The single break model to the X-ray and
$\gamma$-ray data, which could be
applicable if a separate electron population produces the radio
emission, results in a comparable magnetic field strength (4.6$\times 10^{-6}$
G) and cooling break energy (5.5$\times 10^{10}$ keV).

Horns et al. (2006) 
propose a hadronic model for the $\gamma$-ray emission from the Vela X
cocoon, wherein the emission is the result of the decay of neutral 
pions produced in proton-proton collisions in the cocoon.
This model 
requires a
number density 
of $n > 0.6 {\rm\ cm}^{-3}$ for the target material. 
However, the density of thermal emission from our observations is 
only $\sim$0.09 cm$^{-3}$, 
which is too low to accommodate the 
hadronic scenario. Moreover, we have considered an electron/ion number
density ratio consistent with cosmic abundances; the presence of
ejecta will increase this ratio, further reducing the derived gas
density. Our observation only covers the southernmost portion of
the cocoon, so perhaps a higher density is present along the northern
parts. More X-ray observations of Vela X are needed to determine whether or not
the $\gamma$-ray emission from hadronic processes are feasible for other regions
of the PWN.

\acknowledgments{PO Slane acknowledges support
from NASA contract NAS8-03060.}

\clearpage

\begin{table}
\caption{Best-fit Values}
\begin{center}
\begin{tabular}{lcc}
\hline
\noalign{\smallskip}
Parameter &  Value using Background 1 & Value using Background 2 \\
\hline\noalign{\smallskip}
 $N_H$ ($\times 10^{20}$) &  1.6$^{+0.3}_{-0.2}$ &
1.6$^{+0.5}_{-0.5}$ \\ 
 kT (keV) &  0.48$^{+0.05}_{-0.06}$ &  0.50$^{+0.07}_{-0.06}$ \\
 $$[O]$^{1}$ & 1.7$^{+0.1}_{-0.1}$ & 1.6$^{+0.2}_{-0.1}$ \\
 $$[Ne]$^{1}$ & 5.4$^{+0.4}_{-0.4}$ & 4.7$^{+0.4}_{-0.4}$\\
 $$[Mg]$^{1}$ & 3.3$^{+0.6}_{-0.3}$ & 2.9$^{+0.8}_{-0.7}$\\
 $$[Fe]$^{1}$ & 1.4$^{+1.0}_{-0.3}$ & \\
 $\tau$ (10$^{10}$s/cm$^3$) &  4.8$^{+1.7}_{-0.8}$ & 4.0$^{+1.2}_{-0.9}$\\
 $\Gamma$ &  2.3 $\pm0.04$ & 2.2$\pm0.1$ \\ 
 Flux (total)$^{2}$ & 2.5$\pm0.4$ & 4.0$\pm0.4$ \\
 Flux (unabsorbed)$^{2}$ &  2.7$\pm0.4$  & 4.2$\pm0.4$\\
 Flux (power law)$^{2}$ & 1.5$\pm0.2$ & 2.5$\pm0.2$\\
\hline
\hline
\multicolumn{3}{l}{$^{1}$ Abundances relative to solar.}\\
\multicolumn{3}{l}{$^{2}$ Flux in units of 10$^{-11}$ erg
  cm$^{-2}$s$^{-1}$ and averaged among the 3 detectors.}\\
\multicolumn{3}{l}{Background 1 is within the PWN.}\\
\multicolumn{3}{l}{Background 2 is outside the PWN.}\\
\end{tabular}
\end{center}
\end{table}

\begin{table}
\caption{Synchrotron + Inverse Compton Scattering Model Values}
\begin{center}
\begin{tabular}{lc}
\hline
\noalign{\smallskip}
Parameter &  Value \\
\hline
\noalign{\smallskip}
A$_{1}$ (particles/erg/cm$^3$ (s/g/cm)$^{-\sigma}$) & 2$\times10^{-29}$  \\
E$_{1}$ (keV)    & 5$\times10^{6}$    \\
E$_{2}$ (keV)    & 5.5$\times10^{10}$ \\
B (G)            & 5$\times10^{-6}$   \\
$\sigma_{1}$     & 1.75 \\ 
$\sigma_{2}$     & 2.4 \\
$\sigma_{3}$     & 4.2 \\
\hline
\hline
\end{tabular}
\end{center}
\end{table}

\clearpage

\end{document}